\documentclass[sigconf]{acmart}

\usepackage{graphicx} 
\usepackage{natbib}  
\usepackage{caption} 
\usepackage[switch]{lineno}

\usepackage{url}            
\usepackage{booktabs}       
\usepackage{amsfonts}       
\usepackage{nicefrac}       
\usepackage{threeparttable}
\usepackage{multirow}
\usepackage{graphicx}
\usepackage{amsmath}
\usepackage{color}
\usepackage{subfigure}
\usepackage{makecell}
\usepackage{multicol}
\usepackage{listings}
\usepackage{wrapfig}

\usepackage[linesnumbered, ruled]{algorithm2e} 
\usepackage{algorithmic}

\usepackage{tikz}

\newcommand{\AlgName}{\textsc{DeepSweep}\xspace}

\copyrightyear{2021}
\acmYear{2021}
\setcopyright{acmcopyright}
\acmConference[ASIA CCS '21] {Proceedings of the 2021 ACM Asia Conference on Computer and Communications Security}{June 7--11, 2021}{Virtual Event, Hong Kong.}
\acmBooktitle{Proceedings of the 2021 ACM Asia Conference on Computer and Communications Security (ASIA CCS '21), June 7--11, 2021, Virtual Event, Hong Kong}
\acmPrice{15.00}
\acmISBN{978-1-4503-8287-8/21/06}
\acmDOI{10.1145/3433210.3453108}

\newenvironment{packeditemize}{
\begin{list}{$\bullet$}{
\setlength{\labelwidth}{8pt}
\setlength{\itemsep}{0pt}
\setlength{\leftmargin}{\labelwidth}
\addtolength{\leftmargin}{\labelsep}
\setlength{\parindent}{0pt}
\setlength{\listparindent}{\parindent}
\setlength{\parsep}{0pt}
\setlength{\topsep}{3pt}}}{\end{list}}

\begin{document}
\fancyhead{}

\title{\AlgName: An Evaluation Framework for Mitigating DNN Backdoor Attacks using Data Augmentation}

\author{Han Qiu}
\affiliation{%
  \institution{Tsinghua University}
  \country{China}}
  \email{qiuhan@tsinghua.edu.cn}

\author{Yi Zeng}
\affiliation{%
  \institution{University of California San Diego}
  \country{USA}}
\email{y4zeng@eng.ucsd.edu}

\author{Shangwei Guo}
\affiliation{%
 \institution{Chongqing University}
 \country{China}}
 \email{swguo@cqu.edu.cn}

\author{Tianwei Zhang}
\authornote{Tianwei Zhang is the corresponding author.}
\affiliation{%
 \institution{Nanyang Technological University}
 \country{Singapore}}
 \email{tianwei.zhang@ntu.edu.sg}

\author{Meikang Qiu}
\affiliation{%
 \institution{Texas A\&M University Commerce}
 \country{USA}}
 \email{meikang.qiu@tamuc.edu}

\author{Bhavani Thuraisingham}
\affiliation{%
 \institution{The University of Texas at Dallas}
 \country{USA}}
 \email{bhavani.thuraisingham@utdallas.edu}
\renewcommand{\shortauthors}{}

\renewcommand{\shortauthors}{Trovato and Tobin, et al.}

\begin{abstract}
Public resources and services (e.g., datasets, training platforms, pre-trained models) have been widely adopted to ease the development of Deep Learning-based applications. However, if the third-party providers are untrusted, they can inject poisoned samples into the datasets or embed backdoors in those models. Such an integrity breach can cause severe consequences, especially in safety- and security-critical applications. Various backdoor attack techniques have been proposed for higher effectiveness and stealthiness. Unfortunately, existing defense solutions are not practical to thwart those attacks in a comprehensive way.

In this paper, we investigate the effectiveness of data augmentation techniques in mitigating backdoor attacks and enhancing DL models' robustness. An evaluation framework is introduced to achieve this goal. Specifically, we consider a unified defense solution, which (1) adopts a data augmentation policy to fine-tune the infected model and eliminate the effects of the embedded backdoor; (2) uses another augmentation policy to preprocess input samples and invalidate the triggers during inference. We propose a systematic approach to discover the optimal policies for defending against different backdoor attacks by comprehensively evaluating 71 state-of-the-art data augmentation functions. Extensive experiments show that our identified policy can effectively mitigate eight different kinds of backdoor attacks and outperform five existing defense methods. We envision this framework can be a good benchmark tool to advance future DNN backdoor studies.
\end{abstract}



\maketitle

\section{Introduction}

The past several years have witnessed the rapid development of Deep Learning (DL) technology. Various DL models today are widely adopted in many scenarios, e.g., image classification~\cite{chan2015pcanet,targ2016resnet}, speech recognition~\cite{deng2014ensemble}, language processing~\cite{collobert2008unified,lample2016neural}, robotics control \cite{lillicrap2015continuous,zhu2017target}. These applications significantly enhance the quality of life. With the increased complexity of Artificial Intelligence tasks, more sophisticated DL models need to be trained, which require large-scale datasets and huge amounts of computing resources. 

To reduce the training cost and effort, it is now common for developers to leverage third-party resources and services for efficient model training. Developers can download state-of-the-art models from the public model zoos or purchase them from model vendors. They can also download or purchase valuable datasets from third parties and train the models by themselves. A more convenient way is to utilize public cloud services (e.g., Amazon SageMaker~\cite{liberty2020elastic}, GoogleVision AI~\cite{hosseini2017google}, Microsoft Computer Vision~\cite{han2013enhanced}, etc.), which can automatically deploy the training environment and allocate hardware resources based on users' demands. 

However, new security threats are introduced to DNN models when the third party is not trusted. One of the most severe threats is the DNN backdoor attacks~\cite{li2020backdoor}: the adversary injects a backdoor into the victim model, causing it to behave normally over benign samples, but predict the samples with an attacker-specified trigger as wrong labels desired by the adversary. Typically, a backdoor injection can be achieved by directly modifying the neurons~\cite{liu2017trojaning} or poisoning the training datasets~\cite{gu2017badnets}. In practice, the developer may obtain a poisoned dataset if the source is untrusted.  It is hard to detect such a threat as a very small ratio of malicious samples are sufficient to generate a backdoor model. When the developer outsources the model training task to an untrusted cloud provider, the adversary can inject the backdoor by either dataset poisoning or parameter modifications. It is then difficult to detect the existence of backdoors, as the model only has anomalous predictions on samples with triggers, which are agnostic to the developer. 

It is important to have an effective method to address these severe threats. Past works proposed some approaches to detect the existence of backdoors or eliminate them from the infected models. Unfortunately, most of them are not comprehensive enough to cover different types of attack techniques and trigger patterns. For instance, \cite{wang2019neural} is effective against the single-target attack but fails to identify the all-to-all attack where there are more than one target label for the malicious samples~\cite{gu2017badnets}. \cite{liu2019abs} cannot defeat attacks with complex triggers (e.g., global patterns), as claimed in that paper. More importantly, most of these defense works only consider traditional pattern-based attacks, while ignoring the recently discovered advanced ones (e.g., invisible attacks \cite{li2019invisible}). Detailed discussions about these prior works can be found in Section \ref{sec:pastwork}, and some of them are evaluated as baselines in Section \ref{sec:comare-other}.

We argue that it is extremely difficult to design a comprehensive defense method for various backdoor attacks, especially for unknown ones. The rationale behind this argument is that backdoor attacks have no standardization or restrictions over the design of trigger patterns. Different from Adversarial Examples (AE) where the adversarial perturbation is strictly bounded, a backdoor attacker can inject a trigger with an arbitrary size, location, and content to the samples. This incurs insurmountable challenges for the defender to consider and cover all possible trigger and backdoor designs. Hence, instead of building an omnipotent defense, we wonder \emph{if it is possible to have a system or method, which is able to automatically produce solutions to mitigate backdoor attacks within known categories?} With such a system, developers can quickly acquire a new defense, when new attacks are introduced. 

To achieve this goal, we design \AlgName, a first-of-its-kind framework for systematic evaluations of DNN backdoor attacks. \AlgName leverages \emph{data augmentation} to protect DNN models. Data augmentation \cite{shorten2019survey} adopts various image transformations to enrich the datasets. It has become a common technique to enhance model performance and generalization. Recently researchers repurposed it to secure machine learning systems against AEs \cite{raff2019barrage,zeng2020data,qiu2020mitigating}. Since DNN AEs share similar features with backdoor attacks \cite{pang2020tale,jin2020unified}, we propose to use data augmentation techniques for backdoor defense. \cite{li2020rethinking} made an initial attempt by preprocessing trigger-patched samples with simple augmentations. These transformation functions can defeat backdoor attacks with simple trigger patterns, but become ineffective against advanced attacks.  

A successful backdoor attack relies on both the backdoor embedded in the infected model and triggers in the malicious samples. Hence, \AlgName introduces a new backdoor-aware DL pipeline, which integrates \emph{model fine-tuning} and \emph{input preprocessing} with data augmentation. Given an infected model\footnote{If the defender is only given a poisoned dataset, he can first train an infected model and then follow the next two steps. For simplicity, we only consider the case that a compromised model is given throughout the paper}, this pipeline consists of two phases. During the fine-tuning phase, \AlgName adopts an augmentation policy to preprocess clean samples which are further used to retrain the model for a few epochs. This fine-tuning phase is able to alter the model decision boundaries and break the backdoor impact. 
During the inference phase, each sample (either clean one or trigger-patched one) is first preprocessed by another transformation policy before prediction by the fine-tuned model. This phase aims to perturb the trigger patterns. The combination of these two steps can break the connection between the backdoor in the model and the corresponding trigger in the sample. 

The core of the pipeline is the two augmentation policies. They must be able to correct the labels of malicious samples while maintaining high performance for normal data. \AlgName performs a comprehensive study to evaluate and discover the qualified policies. Specifically, \AlgName is equipped with a backdoor database, which contains representative attack instances from known categories. It also includes a data augmentation library of common image transformation functions. Here we must notice the difference between policy and function: we can have a policy composed of multiple functions. We devise a systematic approach to heuristically search and identify the optimal functions and their combinations, which can be effectively used in the pipeline to mitigate any attacks within the considered categories.  

The significance of \AlgName is twofold. First, we use it to discover a unified defense solution to mitigate backdoor threats. Six augmentation functions are shortlisted from 71 functions to form two transformation policies used for fine-tuning the model and preprocessing the inference samples. Evaluations indicate that this lightweight solution can significantly reduce the success rates of 8 common backdoor attacks, covering different techniques (BadNet \cite{gu2017badnets}, Neural Trojan \cite{liu2017trojaning}, invisible backdoor \cite{li2019invisible}), trigger patterns (square, watermark, adversarial perturbation), attack modes (single-target, all-to-all), datasets (Cifar10, GTSRB, PubFig) and models (ResNet-18, LeNet-8, VGG-16). It can also outperform five state-of-the-art works (Neural Cleanse \cite{wang2019neural}, Fine-pruning and Fine-pruning with Fine-tuning \cite{liu2018fine}, FLIP \cite{li2020rethinking}, and ShrinkPad-4 \cite{li2020rethinking}). 

Second, our framework and method are extensible. New attacks and data augmentation functions can be easily integrated into \AlgName for evaluation. 
Although analysis and evaluation frameworks for adversarial examples have been introduced \cite{papernot2016cleverhans,ling2019deepsec,rauber2017foolbox,nicolae2018adversarial}, to the best of our knowledge, there is still a lack of similar platforms for comprehensive evaluation and analysis of DNN backdoor attacks. We expect \AlgName to be such a  valuable framework for researchers and practitioners to understand the mechanisms of backdoor threats, and to build more efficient and effective defenses for robustness enhancement of DNN models. We opensource the \AlgName and welcome the public to contribute to its future development\footnote{https://github.com/YiZeng623/DeepSweep}. 
The key contributions of this paper are:

\begin{packeditemize}
    \item We design a new framework, which is able to automatically evaluate and generate defense methods against backdoor attacks; 
    \item We identify an end-to-end solution based on data augmentation techniques to remove the backdoor via fine-tuning and compromise the trigger effects via inference preprocessing;
    \item We conduct extensive experiments to show our approach is comprehensive and general against different types of attacks, and outperform other state-of-the-art defenses.
\end{packeditemize}

The rest of this paper is organized as follows. Section \ref{sec:bg-attack} introduces the background of backdoor attacks, followed by the analysis of existing defenses in Section \ref{sec:bg-defense}. Section \ref{sec:framework} describes the design of our \AlgName framework. We present one unified solution identified from this framework in Section \ref{sec:solution} and extensive evaluation in Section \ref{sec:eval}. We discuss in Section \ref{sec:discussion} and conclude in Section \ref{sec:conclusion}.

\section{Preliminaries of Backdoor Attacks}
\label{sec:bg-attack}
In a backdoor attack, the adversary attempts to tamper with the integrity of the victim model. The compromised model still has state-of-the-art performance for normal samples. However, for an input sample containing the trigger, the model will predict a wrong label, which can be pre-determined by the attacker, or an arbitrary unmatched one. Formally, given a DNN model $f_\theta$ with parameters $\theta$, a backdoor attack can be formulated as a tuple $(\Delta\theta, \delta)$, where $\Delta\theta$ is the backdoor injected by the adversary to the model parameters, and $\delta$ is an attacker-specified trigger. Then the backdoor model $f_{\theta+\Delta\theta}$ exhibits the following behaviors for normal samples and trigger-patched samples, respectively:

\vspace{-1em}
\begin{align}\label{equ:backdoor}
    & f_{\theta+\Delta\theta}(x)=f_{\theta}(x),\forall x\in \mathcal{X},\\\label{equ:behavior}
    & f_{\theta+\Delta\theta}(x+\delta)\neq f_{\theta+\Delta\theta}(x),\forall x\in \mathcal{X},
\end{align}
\vspace{-2em}

\subsection{Embedding Techniques}
The adversary has multiple ways to embed the backdoor into the DNN model during either the training or deployment phase.

\noindent\textbf{Data poisoning}. 
This applies to the scenario where the developer trains a model based on an untrusted dataset \cite{gu2017badnets,chen2017targeted}. To poison a dataset, the adversary picks some training samples, tampers with a certain portion of each sample with a trigger pattern, assigns them the desired labels different from the correct ones, and then incorporates them into the training set. The model trained from this poisoned set will recognize such a relationship between the trigger and the assigned labels. During the inference phase, it predicts wrong labels whenever the input samples contain such a trigger. 

\noindent\textbf{Parameter modification}. 
This occurs when the adversary has access to a well-trained clean model. Instead of poisoning the training dataset, he can directly modify some critical parameters to make the model malfunction \cite{liu2017trojaning}. Specifically, the adversary investigates the neurons in the model and selects some which are substantially susceptible to the input variations. Then he designs a trigger pattern that can cause these selected neurons to achieve large activation values. By fine-tuning the model with such patterns, those critical neuron values are modified to recognize the triggers. 

\noindent\textbf{Transfer learning}. 
In addition to directly compromising the model or training set, the backdoor can also be propagated via transfer learning. A teacher model can transfer the knowledge and recognition capability to the student models via fine-tuning. Past works discovered that it is also possible to transfer the backdoor from the teacher model to the student model \cite{yao2019latent,WaNe20transfer}. Hence, an adversary can train a backdoor teacher model and make it available in public platforms or model zoos. 
Then, users download this model and perform transfer learning to train a new model, which can inherit the vulnerability, even the student model is fine-tuned with a clean dataset for a totally different task.

\subsection{Trigger Designs}
There are a variety of designs for the malicious triggers in the inference sample to activate the backdoor. These designs can be used to categorize the backdoor attacks. 
Here, we category the trigger designs as the following four patterns. 

\noindent\textbf{Local patterns}.
The most common option is to modify a small block with several pixels at the corner of the image. For instance, \cite{gu2017badnets} added a white square onto the right bottom of the image as the trigger. \cite{liu2017trojaning} introduced a colored square to activate the backdoor. Since these patterns are generally tiny and placed at the corner, their existence will not affect the main content of the image, although they are still perceptible. 

\noindent\textbf{Global patterns}.
Different from the local patterns, this type of triggers are usually across the entire image. With large sizes, they are designed to be dim in the background. For instance, watermarks are embedded over the background of the samples \cite{liu2017trojaning}. Chen et al. \cite{chen2017targeted} proposed to blend a large trigger pattern into the original input.

\noindent\textbf{Invisible perturbation}.
Inspired by adversarial examples, invisible triggers are introduced, which are imperceptible perturbations and visually indistinguishable from normal samples. For instance, Li et al. \cite{li2019invisible} regularized the $L_p$-norm of the perturbation to restrict the scale of the trigger. Liao et al. \cite{liao2018backdoor} leveraged the universal adversarial attack technique to generate triggers bounded by the $L_2$ norm. These triggers can make the backdoor attacks stealthier, and it is hard to detect poisoned data from the training set. 

\noindent\textbf{Semantic patterns}.
The above triggers do not have semantic meanings. Researchers also leveraged the semantic component of an image as triggers, such that the trigger-patched samples look very natural. For instance, Chen et al. \cite{chen2017targeted} designed a special pair of glasses as a trigger when it is worn by a person. Bagdasaryan et al. \cite{bagdasaryan2020blind} adopted certain existing features, e.g., green cars or cars with racing stripes to activate the backdoor in the infected model. This does not require modifying the images. Since this type of triggers are fundamentally different from the above ones, they are out of the defense scope of our framework, as discussed in Section \ref{sec:discuss-comp}.

\section{Defense Analysis}
\label{sec:bg-defense}

\subsection{Threat Model and Defense Requirements}
\label{sec:def-req}

We follow the standard threat model of backdoor attacks: the defender obtains a compromised DNN model containing a backdoor from untrusted third parties or trains a DNN model from a poisoned dataset. He deploys the model into a Deep Learning application or service. During inference, the adversary may query the model with malicious samples containing the trigger to activate the backdoor, making the application give wrong predictions. 
The defender aims to invalidate the backdoor from the compromised model. To achieve this goal, a good solution must have the following properties:

\begin{packeditemize}
\item \emph{Robust}: the solution is capable of eliminating the backdoor effectively with a low attack success rate. It should be hard to be evaded even if the adversary knows the defense mechanism.

\item \emph{Comprehensive}: the defense solution is able to cover different types of backdoor attacks, regardless of the size, complexity, and visibility of triggers, as well as the attacker's target labels. 

\item \emph{Functionality-preserving}: this solution has a small impact on the model performance of clean samples. 

\item \emph{Lightweight}: the defender can defeat backdoor attacks efficiently. Given a suspicious model, the defense cost should be much smaller than training a clean model from scratch. During inference, the prediction process cannot incur high overhead either.

\end{packeditemize}

\subsection{Review of Existing Solutions}
\label{sec:pastwork}
Various defense techniques against backdoor attacks have been proposed. We classify them into different categories and check their satisfaction with the above requirements.

\noindent{\textbf{Backdoor detection.}}
The most popular direction is to check if one DL model has an injected backdoor. 
\cite{wang2019neural} adopted boundary outlier detection to identify anomalous models. Some works followed the similar idea to detect the existence of backdoors and utilized different techniques to recover the trigger, such as Generative Adversarial Networks \cite{chen2019deepinspect}, new regularization terms \cite{guo2019tabor}, Generative Distribution Modeling \cite{qiao2019defending}, and Artificial Brain Stimulation \cite{liu2019abs}.

These approaches make two unrealistic assumptions. First, they assume there is only one target label for all malicious samples (i.e., single-target attack). The detection becomes infeasible when the adversary assigns more than one target label to different samples (e.g., all-to-all attack \cite{gu2017badnets}). Second, they assume the trigger has a small size and simple pattern. Complex triggers such as global patterns can invalidate these approaches. Hence, these solutions cannot meet the \emph{comprehensiveness} requirement.

\cite{xu2019detecting} proposed another detection approach without the above assumptions. It builds a classifier to distinguish benign and infected models. To have higher coverage and accuracy, it needs to mimic all possible backdoor attacks, which is costly and impractical as there are too many possible ways to perform backdoor attacks against DL models. This solution is thus not \emph{lightweight}.

\noindent{\textbf{Backdoor invalidation.}}
This direction is to directly remove the potential backdoor from the model without detection. \cite{liu2018fine} proposed to use fine-pruning and fine-tuning to break the backdoor effects. 
However, this solution may reduce the prediction accuracy over clean samples, which is not \emph{functionality-preserving}.

\noindent{\textbf{Trigger detection.}}
Instead of checking the suspicious model, this direction focuses on the samples with triggers. It can be applied to two cases. The first case is to detect if the training set contains poisoned samples: \cite{chen2018detecting} discovered that normal and poisoned data yield different features in the last hidden layer's activations; \cite{tran2018spectral} proposed a new representation to classify benign and malicious samples; \cite{du2019robust} adopted differential privacy to detect abnormal training samples. These solutions cannot work when the defender only has the infected model rather than the poisoned data samples, especially when the backdoor is injected via direct neuron modification \cite{liu2017trojaning}. They cannot achieve \emph{comprehensiveness}.

The second case is the online detection of triggers during inference. \cite{gao2019strip} proposed to superimpose a target sample with a benign one from a different class. The prediction result of a benign sample will be altered while a malicious sample will still keep the same due to the triggers. However, this approach may not be \emph{robust} when the superimposed benign image has overlap with the trigger. \cite{chou2018sentinet} proposed to use image processing techniques (e.g., Grad-CAM) to visualize and reveal the trigger. This approach is not \emph{comprehensive} as it requires the defender to know exactly the trigger patterns. 

\noindent{\textbf{Trigger invalidation.}}
The last direction is to directly invalidate the effects of the triggers from the inference samples. \cite{li2020rethinking} proposed to adopt common image transformation operations to preprocess input such that the backdoor model will give correct results for both benign and malicious samples. However, since backdoor models and triggers have very high robustness, this solution is not \emph{comprehensive}, as it can only handle simple triggers, but fail to defeat complex ones (e.g., global patterns). 

Our solution also aims to directly prevent backdoor attacks instead of detecting them. Different from the above works, we combine both the directions of \emph{backdoor invalidation} and \emph{trigger invalidation}, to achieve more robust and comprehensive protection. We present our system design in the next section.

\section{Framework Design}\label{sec:framework}

\begin{figure}[!htbp]
  \centering
  \includegraphics[width=\linewidth]{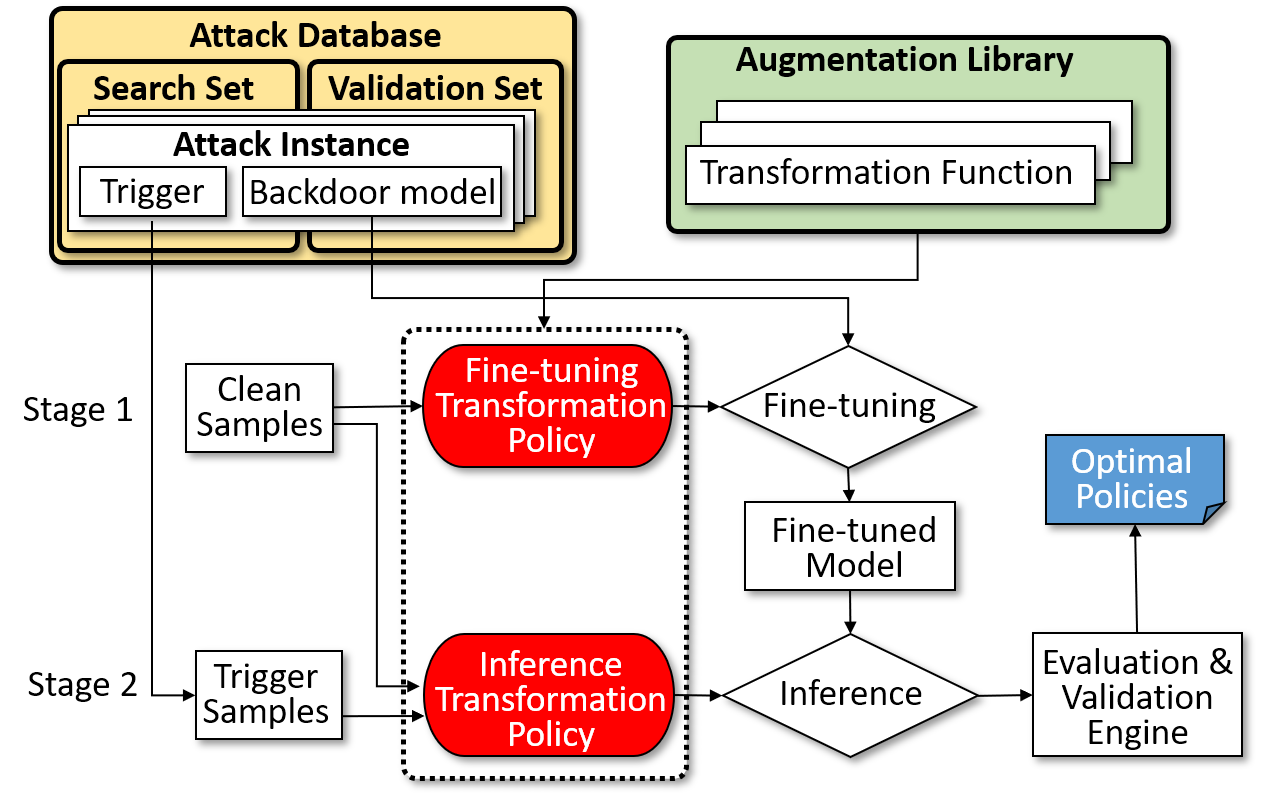}
  \vspace{-2em}
  \caption{Framework overview of \AlgName.}
  \label{fig:pipeline}
  \vspace{-2em}
\end{figure}

\AlgName is designed as a comprehensive framework for evaluating and analyzing the effectiveness of model fine-tuning and input preprocessing in DNN backdoor mitigation. It can help researchers to understand the mechanisms of different backdoor attacks, and design qualified defense solutions. Figure \ref{fig:pipeline} depicts the overview of \AlgName, consisting of an Attack Database, an Augmentation Library, a two-stage pipeline, and an Evaluation \& Validation Engine. The Attack Database and Augmentation Library modules are designed to be extensible, so users can easily incorporate more attacks or functions into consideration. Besides, these modules are also independent of each other, allowing researchers to flexibly adjust their defense strategies (e.g., fine-tuning only or inference preprocessing only). Below we describe each module in detail. 

\subsection{Attack Database}

\begin{figure}[!htbp]
  \centering
  \includegraphics[width=0.999\linewidth]{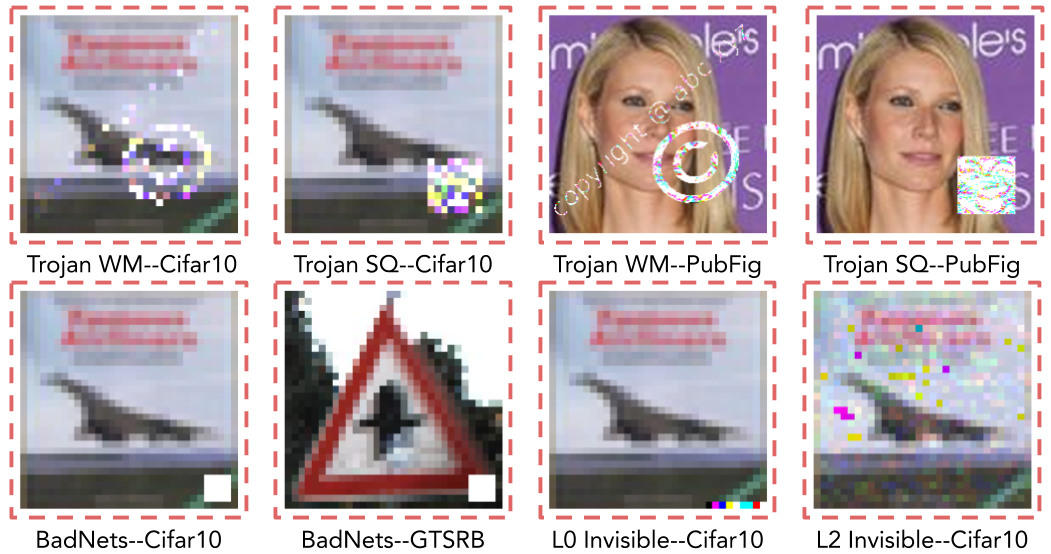}
  \vspace{-2em}
  \caption{Trigger-patched samples in various backdoor attacks in \AlgName.}
  \label{fig:allattack}
    \vspace{-1em}
\end{figure}

This module contains different kinds of state-of-the-art backdoor attacks from past literature for evaluation. Table \ref{tab:setting} summarizes the configurations of these attacks in our current implementation, as well as the target models and datasets. Figure \ref{fig:allattack} shows the trigger-patched samples for each attack instance. We mainly replicate the same implementations as the original papers. 

\begin{table}[!htbp]
\centering
\newcommand{\tabincell}[2]{\begin{tabular}{@{}#1@{}}#2\end{tabular}}
\scalebox{0.75}{
\begin{tabular}{c|c|c|c|c|c}
\Xhline{1pt}
\textbf{Attack} & \textbf{Dataset}  & \textbf{Model} &  \makecell{\textbf{Target} \\\textbf{Label}}  & \makecell{\textbf{Poisoning} \\\textbf{Ratio}} & \textbf{Type}\\
\Xhline{1pt}
\multirow{2}*{Trojan (WM)} & Cifar10 & ResNet-18 & `7'&10\% & Validation\\\cline{2-6}
& PubFig & VGG-16 & `0'&10\% & Search \\ \hline
\multirow{2}*{Trojan (SQ)} & Cifar10 & ResNet-18 & `7'&10\% & Validation\\\cline{2-6}
& PubFig & VGG-16 & `0'&10\% & Validation \\ \hline
BadNets (All-to-all) & Cifar10 & ResNet-18 & `i+1'&10\% & Validation \\ \hline
BadNets (Single target) & GTSRB & LeNet-8 & `33'&10\% & Validation \\ \hline
L2 Invisible & Cifar10 & ResNet-18 & `3'& 5\% & Search \\ \hline
L0 Invisible & Cifar10 & ResNet-18 & `4'& 5\% & Search \\ \hline
\Xhline{1pt}
\end{tabular}}
\caption{Eight kinds of backdoor attacks over three different datasets are collected in \AlgName.}
\label{tab:setting}
  \vspace{-2em}
\end{table}

The first kind of instances is the trojan attack, proposed in \cite{liu2017trojaning}. 
We consider two trigger patterns: a watermark (WM) across the background of the image, and a square-shape (SQ) trigger on the right bottom of the image. 
For each pattern, we consider two datasets including the Cifar10 and PubFig~\cite{kumar2009attribute} datasets. 
Cifar10 is a wildly-adopted dataset for image classification. It contains 50000 training images and 10000 testing images. 
We train a ResNet-18 \cite{he2016deep} backdoor model with the attacker's target label as class `7:Horse'. 
The PubFig dataset contains 11070 training images and 2768 testing images of 83 celebrities. We train a VGG-16 model and set the attacker's label as `0:Adam Sandler'. 
These models are compromised with 10\% of poisoned samples. 

The second type of attack is BadNet~\cite{gu2017badnets}. 
The trigger is a white square of $5\times 5$ pixels located on the right bottom of the image. 
We consider two attack modes: in ``all-to-all'', the target label of a sample from class $i$ is set to be class $i+1$. 
This is realized in the Cifar10 dataset with a poisoning ratio of 10\%. 
In ``single-target'', we use the GTSRB dataset \cite{stallkamp2012man} which contains 35228 training samples and 12630 testing samples in 43 classes.
Here, we directly obtain a backdoor model (LeNet-8) from \cite{wang2019neural}, which has been compromised by the BadNets technique \cite{gu2017badnets}. 
The target label of all trigger samples is set as `33:turn right ahead'.

The third type of instances is invisible attack \cite{li2019invisible}, where the triggers are adversarial perturbations bounded by either L0-norm or L2-norm. 
These attacks are implemented on the Cifar10 dataset, with a poisoning ratio of 5\%, which is large enough to embed the backdoor into the model. The target class is obtained by forward-passing the trigger to a pre-trained clean ResNet-18 model: `3:Cat' for L2 attack and `4:Deer' for L0 attack.

\noindent\textbf{Search and Validation.}
We split these attack instances into two sets. The first set is used to search for the optimal transformation policies, while the second set is used to validate if the searched policies are general for other attacks as well. Specifically, in our current implementation, we choose one instance from each pattern category as the representative in the search set: trojan with WM on PubFig for global pattern triggers, L0 attack for local pattern triggers\footnote{Although L0 attack follows the adversarial example technique, the generated trigger is still visible, located at the right bottom corner (Figure \ref{fig:allattack}). Hence, we classify it as a local pattern backdoor, and select it for policy searching}, and L2 attack for invisible perturbation triggers. The rest five attacks are in the validation set, as shown in Table \ref{tab:setting}. 

\subsection{Augmentation Library}

\AlgName evaluates and selects certain functions from the Augmentation Library to build the backdoor defense. 
In our current implementation, this library includes 71 common image transformation functions. These functions can be classified into four categories. The first three categories contain 65 functions, selected from the Albumentations library \cite{buslaev2020albumentations}. These functions mainly include some basic operations like flip, transpose, Gamma transformation, median filter, Gaussian noise, etc. They are widely used for model generalization enhancement. The last category contains 6 functions from the FenceBox library \cite{qiu2020fencebox}. They are originally adopted to mitigate adversarial examples and improve model robustness. Below we briefly describe these categories, with a detailed list in Table \ref{tab:auglib} in the appendix. 

\noindent\textbf{C1: Affine-Transformation}.
This category includes 22 augmentation functions. They mainly distort the images by significantly changing the pixel locations or dropping a certain ratio of pixels. Since some backdoor attacks inject the triggers by changing selected pixels, these Affine-Transformation-based augmentation functions can potentially drop certain pixels or compromise the patterns, making the trigger unrecognizable by the infected model. 

\noindent\textbf{C2: Compression/Quantization}.
This category contains 16 functions to compress or quantize the images. Some functions follow the standard image compression algorithms to resize the image. Other functions quantize the pixel values to fewer bits. These operations may also introduce perturbations over the trigger patterns. 

\noindent\textbf{C3: Noise Injection/Channel Distortion}.
This category has 27 augmentation functions, which inject random noise or distort different channels of the images. Some operations randomly adjust the attributes (e.g., brightness, contrast) of the images. Some functions randomly drop, shift, or shuffle pixels. There are also some functions to achieve special effects like blur, shadow, rain/snow/fog, etc. These random operations can also bring large perturbations to the images while maintaining their semantics. 

\noindent\textbf{C4: Advanced Transformation}.
This category contains 6 sophisticated transformation functions: Pixel Deflection \cite{prakash2018deflecting}, Bit-depth Reduction \cite{DBLP:conf/ndss/Xu0Q18}, Random Sized Padding Affine~(RSPA) \cite{qiu2020fencebox}, Stochastic Affine Transformation~(SAT) \cite{zeng2020data}, SHIELD \cite{das2018shield}, and Feature Distillation \cite{liu2019feature}. They are initially designed to mitigate adversarial examples. They preprocess the input samples with non-differentiable or non-deterministic operations to obfuscate the gradients, making it difficult or infeasible to generate adversarial perturbations from the original images. Since backdoor attacks and adversarial examples share similar features, we also include these operations to evaluate their effectiveness in backdoor removal.

\subsection{Two-stage Defense Pipeline}
\label{sec:pipeline}
\AlgName establishes a DL pipeline based on data augmentation to protect the models from backdoor attacks. As we discussed in Section \ref{sec:pastwork}, Li et al. \cite{li2020rethinking} also introduced image transformations over the inference samples to remove the triggers. These transformations should be intensive enough to affect the triggers, but also lightweight to maintain the model performance on clean samples. Our evaluations in Section \ref{sec:comare-other} show that this trade-off between security and model usability is difficult to achieve for just preprocessing inference samples.

In contrast, our pipeline consists of two stages, both of which adopt the data augmentation. At stage 1, we introduce a Fine-tuning Transformation Policy, which is an ensemble of certain functions selected from the Augmentation Library. \AlgName applies this policy to a small set of clean samples and then uses the transformed output to fine-tune the infected model for a few epochs. At the end of this stage, we can obtain a fine-tuned model, which has different decision boundaries from the original infected model. Such changes can weaken the effects of the backdoor to some extend. 

At stage 2, we introduce an Inference Transformation Policy, which is another ensemble of functions from the Augmentation Library. This policy is used online to preprocess each inference sample (clean and trigger-patched). The preprocessed images are then fed into the fine-tuned model for prediction. This transformation policy is expected to disturb the triggers and rectify the model output of malicious samples.

The goal of our evaluation framework is to identify functions from the Augmentation Library to form the two policies in the pipeline, that can effectively mitigate the backdoor threats in the Attack Database. Below we design a new approach to systematically discover the optimal solutions.

\subsection{Evaluation \& Validation Engine}
\label{sec:evaMethod}

This module is designed to evaluate the functions from the Augmentation Library, and produce the optimal policies. It contains a set of metrics and a novel evaluation methodology.

\noindent\textbf{Metrics.} The transformation policies in the pipeline need to meet the defense requirements in Section \ref{sec:def-req}. Particularly, it needs to be robust for backdoor elimination. We adopt the Attack Success Rate (ASR) over trigger-patched samples to quantify this property. ASR is calculated as the ratio of those samples that are still predicted as the adversary's desired labels. A lower ASR indicates the higher robustness of the solution. Besides, our policies also need to be functionality-preserving for maintaining model performance. We adopt model accuracy (ACC) over clean samples to measure this requirement. A higher ACC indicates the policies have a smaller impact on model usability. Both ASR and ACC are measured using 200 different trigger-patched (or clean) samples. 

\noindent\textbf{A heuristic search algorithm.}
We introduce an approach to heuristically identify the optimal policies that can meet the defense requirements. Algorithm \ref{algo:AlgoS} illustrates the process. The entire search process consists of two steps. 

\SetKwInput{KwParam}{Parameters}
\begin{algorithm}[]
\small
    \caption{Searching for optimal policies}
    \label{algo:AlgoS}
    \SetNoFillComment
    \KwIn{Augmentation Library: $T$; Search set in Attack Database: $F$}
    \KwOut{Fine-tuning Policy: $P_{f}$; Inference Policy: $P_{i}$}
    \KwParam{ACC threshold: $\epsilon_{acc}$; ASR threshold:$\epsilon_{asr}$\; 
      \qquad \quad \quad \quad \ \,\,\,\# of functions in $P_{f}: $$n$;}

    \BlankLine

     \tcc{Step 1: shortlist functions}
     \BlankLine
     $S$ = $list\{\}$\;
      \For{$t_{i} \in T$}{
        \tcc{$j-th$ backdoor model $m_j$ in Attack Database\; 
        \quad\, 200 clean samples $d_j^{c}$\; 
        \quad\, 200 trigger-patched samples $d_j^{t}$}
          \For{$(m_j,d_j^{c},d_j^{t}) \in F$}{
            $\widehat{d_j^{c}} = t_i(d_j^{c})$\;
            $\widehat{d_j^{t}} = t_i(d_j^{t})$\;
            $acc_j$ = ACC of $m_j$ over $\widehat{d_j^{c}}$\;
            $asr_j$ = ASR of $m_j$ over $\widehat{d_j^{t}}$\;
          }
          \If{$acc_j > \epsilon_{acc}$ \emph{for each} $j$}{
          $S.append(t_{i})$\;
          }
      }
      Sort $S$ from the lowest average ASR to the highest average ASR\;
      
      \BlankLine
      \tcc{Step 2: obtain optimal policies}
      \BlankLine
      $P_f = S[:n]$\;
      \tcc{10000 clean samples $d_j^{f}$}
      \For{$(m_j,d_j^{f}) \in F$}{
            $\widehat{d_j^{f}} = P_f(d_j^{f})$\;
            $\widehat{m_j}$ = Fine-tune $m_j$ over $\widehat{d_j^{f}}$ for 5 epochs\;
            \tcc{Fine-tune here only needs 5 epochs.}
          }
      $S'$ = $list\{\}$\;
      $\widehat{d^t}_{base} = P_f(d_j^{t})$\;
      $avg_{base}$ = (ASR of $\widehat{m_j}$ over $\widehat{d^t}_{base}$ \textbf{for} all $\widehat{m}$)\;
      $SS$ = set of all possible function combinations from $P_f$\;
      \For{$p \in SS$}{
          \For{$(m_j,d_j^{c},d_j^{t}) \in F$}{
            $\widehat{d_j^{t}} = p(d_j^{t})$\;
            $asr_j$ = ASR of $\widehat{m_j}$ over $\widehat{d_j^{t}}$\;
         }
          \If{$avg_{base}(asr_j) - avg_{j}(asr_j) > \epsilon_{asr}$}{
          $S'.append(p)$\;
          }
      }
      $P_i$ = the policy with the smallest ASR in $S'$\;
    \BlankLine
      \Return $P_f, P_i$
        
\end{algorithm}

The first step is to \textbf{shortlist functions} from the augmentation library. We evaluate each transformation function and select the ones based on their ACCs. Specifically, for a function $t_i$, we consider each backdoor attack in the search set, transform the corresponding 200 clean samples $d_j^{c}$ with $t_i$, and measure the model accuracy of the transformed samples $\widehat{d_j^{c}}$. The function $t_i$ is selected when the accuracy over each attack instance is higher than $\epsilon_{acc}$.

The second step is to \textbf{obtain optimal policies} from the shortlisted functions $S$.
This involves two policies to transform clean data for model fine-tuning and inference data for preprocessing. We consider the Fine-tuning Transformation Policy $P_f$ has $n$ functions.
Then $P_f$ is the top-$n$ functions from $S$ with the lowest ASR. Our experiments show that the order of transformation functions in a policy does not significantly impact the policy effects. 
So we can combine these $n$ functions in an arbitrary order to form $P_f$. We use $P_f$ to transform 10000 clean samples to obtain $d_j^{f}$, and use them to fine-tune the backdoor model for 5 epochs to get $\widehat{m_j}$.

Next, we need to build the Inference Transformation Policy $P_i$ from the shortlisted functions. 
We make this policy contain a subset of functions from $P_f$\footnote{We choose $P_i$ to be a subset of $P_f$ in order to make the online inference lightweight. Our evaluations indicate that this can achieve better defense results than using the entire $P_f$ to preprocess the inference samples (Section \ref{sec:remarks}).}. 
We consider all the possible combinations of functions from $P_f$. We ignore the order of these functions in a policy as this does not affect the ACC or ASR based on our empirical experience. Specifically, for each combination, we measure the corresponding ASR of each attack instance. This is achieved by applying the candidate policy over the 200 trigger-patched samples $d_j^{t}$, and measuring the ASR of the fine-tuned model. We regard a policy as qualified if its average ASR is at least $\epsilon_{asr}$ smaller than the average ASR using $P_f$ as the inference policy. The one with the lowest ASR is finally selected as $P_i$.

\noindent\textbf{Validation.}
After identifying the optimal $P_f$ and $P_i$, we deploy them into the pipeline, and use the attack instances from the validation set of the Attack Database to check if they are effective for other unseen attacks as well. If the ASR and ACC are also satisfactory, we will use these two policies as the final solution. Otherwise, we need to repeat the above search procedure. We can adjust the Attack Database by moving the attacks that are not addressed by the previous policies from the search set to the validation set. Then the searched results from the above procedure will be more powerful and comprehensive.

\section{A Defense Solution Discovered by \AlgName}
\label{sec:solution}

We have used \AlgName to discover the qualified fine-tuning and inference transformation policies. In this section, we describe this end-to-end solution and we set $n=6$, $\epsilon_{acc}=0.7$ and $\epsilon_{asr}=0.01$.

\begin{figure*}[!htbp]
  \centering
  \includegraphics[width=\textwidth]{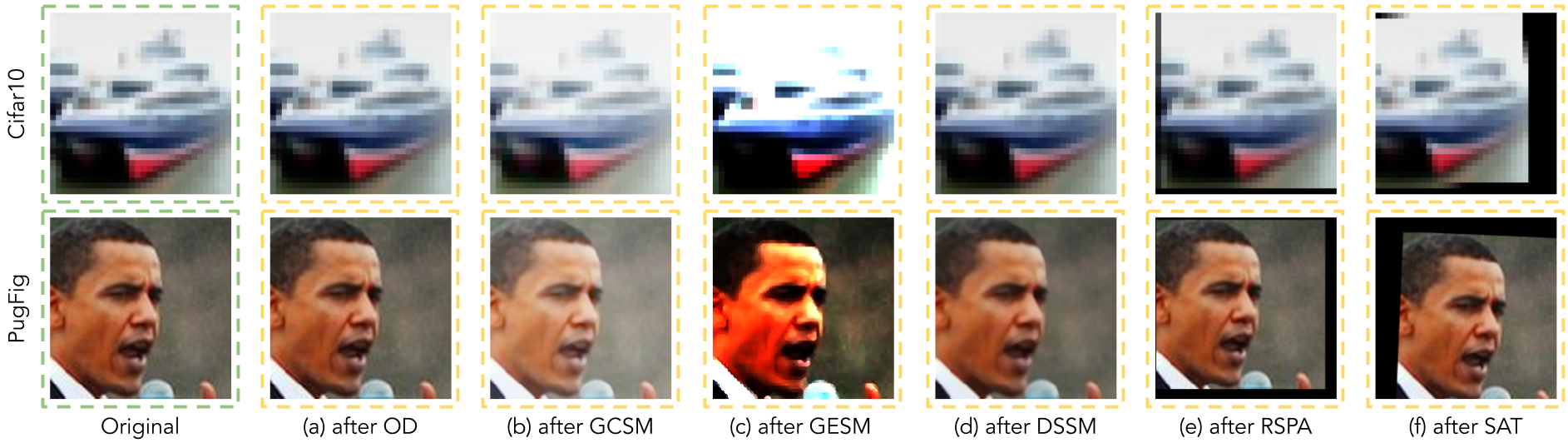}
  \vspace{-2em}
  \caption{Visual results of the transformed images with different augmentation functions individually.}
  \label{fig:candidates}
  \vspace{-1em}
\end{figure*}

\subsection{Shortlisted Augmentation Functions}\label{sec:shortlist}

By scanning the Augmentation Library using Algorithm \ref{algo:AlgoS}, we can acquire a list of defense candidates who satisfy the ACC threshold of 70\%. Table \ref{tab:shortlist} presents the top 6 functions with the lowest average ASR. Figure \ref{fig:candidates} shows the transformed images (one from Cifar10 and one from PubFig) with each augmentation function. Below we describe the basic operation of each function. Detailed algorithms of these functions can be found in the appendix.

\begin{table}[!htbp]
\centering

\newcommand{\tabincell}[2]{\begin{tabular}{@{}#1@{}}#2\end{tabular}}
\scalebox{0.9}{
\begin{tabular}{c|c|p{0.7cm} p{0.7cm}| p{0.7cm} p{0.7cm}|p{0.7cm} p{0.7cm}}
\Xhline{1pt}
\multirow{2}*{Function} & \multirow{2}*{\makecell {\textbf{Average} \\ \textbf{ASR}}}  & \multicolumn{2}{c|}{{Cifar10 (L2)}}  & \multicolumn{2}{c|}{{Cifar10 (L0)}}  & \multicolumn{2}{c}{{PubFig (WM)}} \\

\cline{3-8}
~ &~& ASR & ACC & ASR & ACC & ASR & ACC\\
\Xhline{1pt}

Baseline & \textbf{0.988} & 0.985 & 0.900 & 0.980 & 0.895 & 1.00 & 0.960 \\
\hline 
SAT  & \textbf{0.583} & 0.645 & 0.805 & 0.250 & 0.840 & 0.870 & 0.740 \\
\hline 
GCSM & \textbf{0.595} & 0.680 & 0.790 & 0.285 & 0.815 & 0.820 & 0.940 \\
\hline 
DSSM & \textbf{0.671} & 0.670 & 0.845 & 0.505 & 0.870 & 0.839 & 0.945\\
\hline 
RSPA & \textbf{0.738} & 0.650 & 0.845 & 0.625 & 0.875 & 0.940 & 0.955 \\
\hline 
GESM & \textbf{0.783} & 0.715 & 0.735 & 0.645 & 0.705 & 0.990 & 0.835 \\
\hline 
OD   & \textbf{0.892} & 0.970 & 0.715 & 0.990 & 0.890 & 0.890 & 0.955 \\

\Xhline{1pt}
\end{tabular}}
\caption{Top 6 augmentation functions with $ACC \geq 0.7$.}
\vspace{-2em}
\label{tab:shortlist}
\end{table}

\noindent{\textbf{T1: Optical Distortion (OD).}} 
This function is based on a pincushion distortion~\cite{liu2010pincushion}, which increases the image magnification with the distance from the optical axis. It maps the representation of inputs away from the original one in hyperdimensional space. 
\figurename~\ref{fig:candidates}(a) shows the preprocessed output of two images. We observe lines that do not go through the center of the image are bowed towards the center after this transformation, like a pincushion.

\noindent{\textbf{T2: Median filter with Gamma Compression (GCSM).}} 
A set of median filters are identified to defeat backdoor attacks. 
The first kind of filter is to preprocess the input sample in the gamma space with gamma compression. This gamma compression causes large-value pixels inside the image to bend in together (Figure \ref{fig:candidates}(b)). 
The small-value pixels in the image thus have a better contrast against large-value pixels. Therefore, the median filter can better smoothen those pixels. The default kernel size is $5 \times 5$, and the encoding gamma value is set as 0.6 to lighten the images.

\noindent{\textbf{T3: Median filter with Gamma Extension (GESM).}}
Another type of filter is also performed in the gamma space but with a gamma extension. 
Each pixel is first multiplied by a factor (set as 1.53) to lighten up the images, bend large-value pixels together, and disrupt the continuity between pixels. 
Then, a gamma extension is used to dim the image for obtaining a higher contrast. The image is further scaled down to 75\% of its original size and the same median filter as the first one is conducted in this gamma extension space. 
Such an operation can help the fixed-kernel median filter remove more outliers globally and obtain smoother results. 
Finally, the image is resized to its original size (Figure \ref{fig:candidates}(c)). 

\noindent{\textbf{T4: Median filter with Scaling Down (DSSM).}} 
The third type of median filter works with a scaling down (resize) procedure. The image is first scaled down to 0.8 of its original size. Then the median filter works on the resized figure and finally resizes the smoothened figure back to the original size. This procedure can increase the filter's efficiency when working in the down-scaled space, as neighbor pixel values are merged first during the downscaling. The median filer further reduces the sharpness of the input before resizing it back to the original size. The visual results demonstrate that pixels are indeed smoothened with the help of woring in this downscaled space (see~\figurename~\ref{fig:candidates}(d)).

\noindent{\textbf{T5: Random scaling down with Padding (RSPA).}} ShinkPad \cite{li2020rethinking} has demonstrated the effectiveness of a similar operation at invalidating the BadNets attack \cite{gu2017badnets}. Specifically, the image is first scaled into a smaller size ranging between [0.8-1] of the original one, by dropping random pixels. It is then padded to the original size by randomly choosing a point as the center (Figure \ref{fig:candidates}(e)). 
Such an operation can shift all the pixels away from the actual coordinates. Thus, samples will likely move away from the infected model's original representation output (with a certain accuracy drop). 

\noindent{\textbf{T6: Stochastic Affine Transformation (SAT).}} This preprocessing function is used to distort the image with rotation, scaling, and shifting. SAT first randomly shifts all the pixels horizontally and vertically. Then, it randomly rotates the image to a certain scale. 
Finally, it randomly scales the image up or down to produce the final output. 
The visual effect is shown in Figure \ref{fig:candidates}(f).

\subsection{Optimal Transformation Policies}

Based on the shortlisted augmentation functions, we can now build the transformation policies for model fine-tuning and inference preprocessing. respectively. We use Algorithm \ref{algo:AlgoS} to identify the policies, as described below.

\subsubsection{Fine-tuning Transformation Policy}
\begin{figure}[!htbp]
  \centering
  \includegraphics[width=0.999\linewidth]{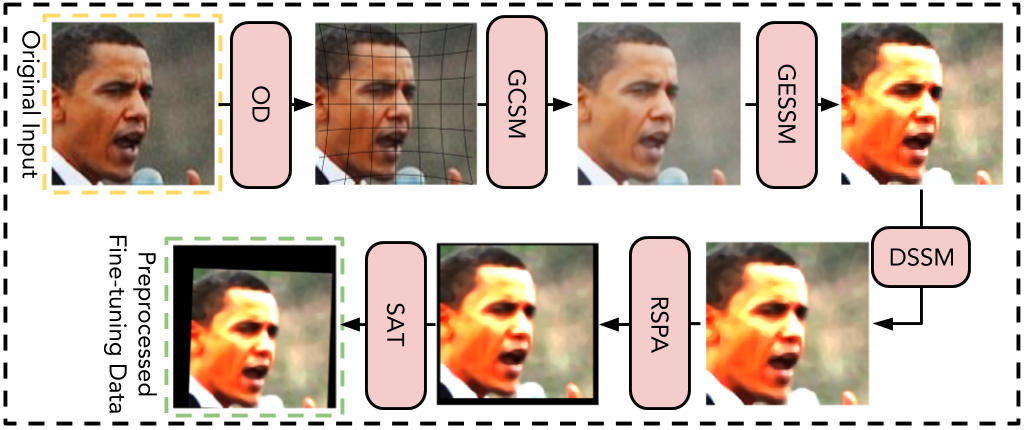}
   \vspace{-2em}
  \caption{Fine-tuning Transformation Policy includes six functions: three affine transformations (OD, RSPA, and SAT) and the three median filters (GCSM, GESM, and DSSM).}
 \vspace{-1em}
  \label{fig:intense}
\end{figure}

This policy includes all the six augmentation functions (T1 -- T6) to preprocess clean samples for fine-tuning. The visual effects are shown in \figurename~\ref{fig:intense}. This policy can introduce significant distortion to the samples. \AlgName only requires a small number of epochs (5 in our experiments) with a few transformed clean samples (10000 for all the models) to fine-tune the model. Then the classification boundaries of the model will be altered against malicious samples patched with the triggers. Besides, the generalization capability of this model is also improved: the model is able to recognize such transformations, and better predict the inference samples preprocessed by the Inference Transformation Policy.

\subsubsection{Inference Transformation Policy}

\begin{figure}[!htbp]
  \centering
  \includegraphics[width=0.999\linewidth]{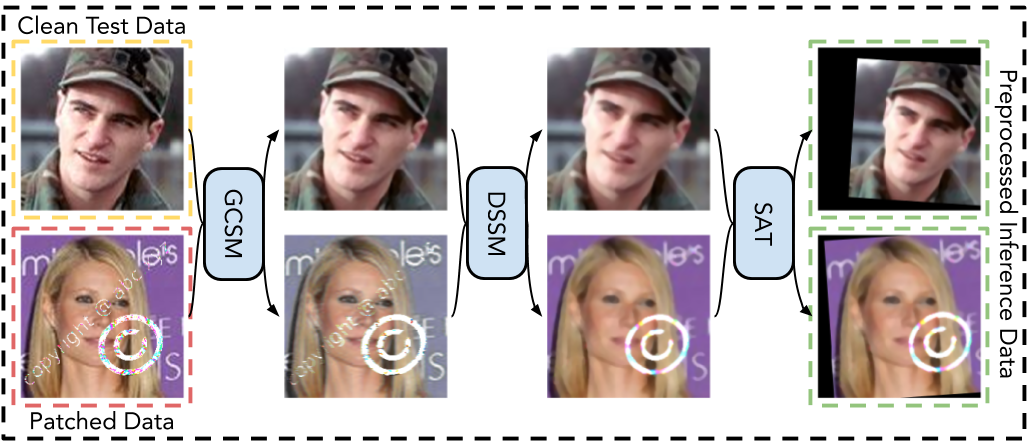}
  \vspace{-2em}
  \caption{Inference Transformation Policy consists of three functions: two median filters affect the triggers from two spaces; SAT helps distort the image. The first row is for a clean image, and the second row is for a patched image.}
  \vspace{-1em}
  \label{fig:light}
\end{figure}

During inference, the transformation policy only includes three operations. The first one is a median filter to smoothen the pixels in the raw input (GCSM). Then a second median filter is integrated with the scaling down mechanism (DDSM). Finally, the Stochastic Affine Transformation (SAT) is adopted over the filtered data to map the pixels away from the original coordinates. This transformation policy is more lightweight than the fine-tuning policy, to achieve better online efficiency. It guarantees the model can predict clean samples correctly while fail to recognize the triggers. Figure \ref{fig:light} shows the results of the inference transformation over clean and trigger-patched samples. 

\section{Evaluation}
\label{sec:eval}
In this section, we conduct extensive evaluations of our identified defense strategy. We demonstrate its effectiveness against attacks in the search set as well as the validation set. We also show its advantages over state-of-the-art defense solutions. We perform model explanations to interpret the effectiveness of this solution. 

We use Keras with Tensorflow backend for the implementations. All the infected models are trained with Adadelta \cite{zeiler2012adadelta} as the optimizer with an initial learning rate of 0.05 for 200 epochs. We conduct the experiments on a server equipped with 8 Intel I7-7700k CPUs and 4 NVIDIA GeForce GTX 1080 Ti GPUs.

\subsection{Effectiveness against Searched Attacks}
\label{sec:remarks}
Table \ref{tab:whyfinetune} shows the evaluation results of the identified policies on the attacks in the search set. It also includes some other strategies based on the two policies. We draw some interesting conclusions from this table. 

\begin{table*}[!htbp]
\centering

\newcommand{\tabincell}[2]{\begin{tabular}{@{}#1@{}}#2\end{tabular}}
\scalebox{0.88}{
\begin{tabular}{c|c|p{1.1cm} p{1.1cm}|p{1.1cm} p{1.1cm}|p{1.1cm} p{1.1cm}|p{1.1cm} p{1.1cm}|p{1.1cm} p{1.1cm}}
\Xhline{1pt}
\multirow{2}*{\textbf{Attack}} & \multirow{2}*{\textbf{Model}}  & \multicolumn{2}{c|}{\textbf{Baseline}} & \multicolumn{2}{c|}{\makecell{\textbf{$P_f$ for Fine-tuning} \\\textbf{$P_i$ for Inference}}} & \multicolumn{2}{c|}{\makecell{\textbf{$P_f$ for Inference}}} & \multicolumn{2}{c}{\makecell{\textbf{$P_f$ for Fine-tuning}}} & \multicolumn{2}{|c}{\makecell{\textbf{$P_f$ for Fine-tuning} \\ \textbf{$P_f$ for Inference}}} \\
\cline{3-12}
~ &~& ACC & ASR  & ACC & ASR & ACC & ASR & ACC & ASR & ACC & ASR\\
\Xhline{1pt}
L2 invisible & \multirow{2}*{ResNet-18 (Cifar10)} & 0.900 & 0.985 & \textbf{0.810} & \textbf{0.180} & 0.610 & 0.420 & 0.785 & 0.390 & 0.790 & 0.205 \\
\cline{1-1}\cline{3-12} 
 L0 invisible & & 0.895 & 0.990 & \textbf{0.825} & \textbf{0.080} & 0.645 & 0.135 & 0.800 & 0.110 & 0.805 & 0.080 \\
\hline 
Trojan (WM)& \multirow{1}*{VGG-16 (PubFig)} & 0.960 & 1.000 & \textbf{0.910} & 0.010 & 0.400 & 0.360 & 0.900 & \textbf{0.000} & 0.840 & 0.010 \\
\Xhline{1pt}
\end{tabular}}
\caption{Evaluation of ACC and ASR with different strategies for attacks in the search set.}
  \vspace{-2em}
\label{tab:whyfinetune}
\end{table*}

First, compared to the baseline where no defense is applied, our solution ($P_f$ for Fine-tuning, $P_i$ for Inference) can indeed mitigate the three backdoor attacks in the search set. The ASR can be kept to be very small values, while the accuracy penalty is acceptable. 

Second, only preprocessing the inference samples (as proposed in \cite{li2020rethinking}) is not effective enough to defeat backdoor attacks. As shown in Table \ref{tab:whyfinetune} ($P_f$ for Inference), inference transformation with the six shortlisted functions can only reduce the ASR of L0 invisible attack to a satisfactory scale. The ASR of L2 invisible attack and Trojan attack with watermarks are still high. More importantly, the model accuracy drops significantly due to the intensive preprocessing of inference samples. This highlights the importance of fine-tuning transformation, which allows the model to recognize such data augmentation operations. 

Third, we consider a strategy that just fine-tunes the model with data augmentation. Similar ideas have been proposed in \cite{liu2018fine}. In our experiment, we only use $P_f$ to transform the clean images and fine-tune the model for a few epochs. The results are shown in the column of `$P_f$ for Fine-tuning' in Table \ref{tab:whyfinetune}. We observe that this strategy can reduce the ASR of these attacks to some extent. However, it is still worse than our optimal solution for both ACC and ASR. One exception is the Trojan attack (WM), where the ASR of this strategy is zero. Unfortunately, this suffers from low generalizability: in the case of the Trojan attack (SQ) on the same dataset, the ASR can reach 100\% (not shown in this table). This indicates the necessity of transformation during inference. 

Fourth, we consider a strategy where the policy $P_f$ is applied to both the fine-tuning and inference stages (the last column in Table \ref{tab:whyfinetune}). Surprisingly, we find the defense results are slightly worse than using $P_i$ for inference transformation and $P_f$ for fine-tuning. This indicates that the fine-tuning and inference policies are not necessarily the same. Using a lightweight transformation policy can reduce the inference overhead, and possibly improve the model performance as well as robustness.

\subsection{Effectiveness against Validation Attacks}
\label{sec:eval-validate}
\begin{figure}[!htbp]
  \centering
  \includegraphics[width=0.99\linewidth]{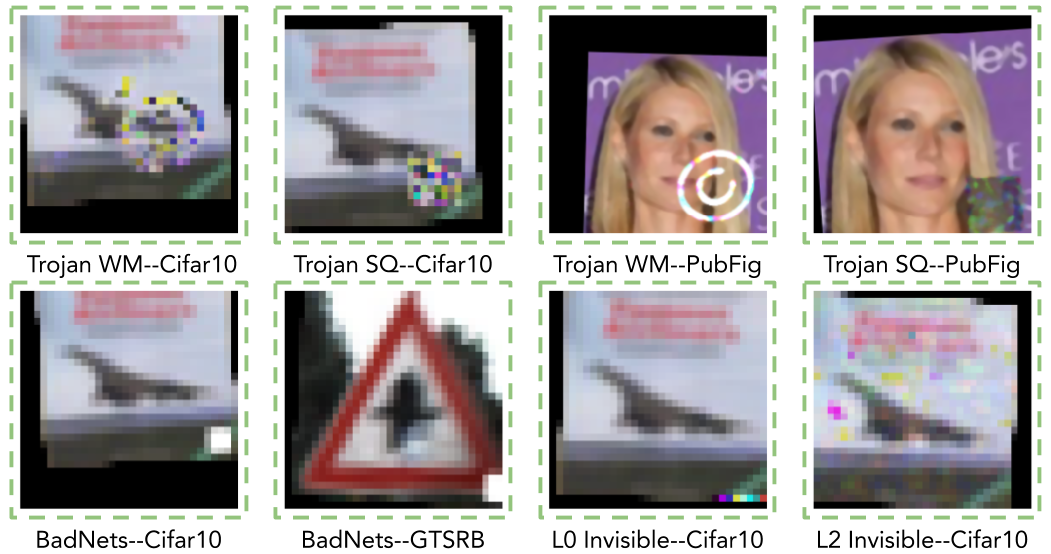}
  \vspace{-1em}
  \caption{Visual results over all attack instances using the Inference Transformation Policy.}
\vspace{-1em}
  \label{fig:defense}
\end{figure}

As discussed in Section \ref{sec:framework}, we only use three attack instances from the search set to discover the optimal policies, which can effectively mitigate all three attacks. Here we show that this solution is general and can mitigate other attacks in the validation set as well. Figure \ref{fig:defense} illustrates the visual results of transformed trigger-patched images for each attack instance, compared to the original ones in Figure \ref{fig:allattack}.

\tablename~\ref{tab:whyfinetune2} shows the ACC and ASR of the target model without and with our transformation policies. We can observe that this solution is still very effective against those attacks while maintaining acceptable model performance. We also measure the strategy with $P_f$ for both fine-tuning and inference transformations, which has slightly worse results. This matches the conclusion in \tablename~\ref{tab:whyfinetune}. 

In summary, \AlgName is able to produce general solutions that are not limited to the attacks used for search evaluation, but also unseen attacks within the same categories of trigger patterns. This proves \AlgName is \emph{comprehensive}. The searched policies can guarantee \emph{robustness} and \emph{functionality-preserving}. The offline fine-tuning only needs 5 epochs, and the online inference contains only simple transformation, making our solution \emph{lightweight}.

\begin{table*}[!htbp]
\centering

\newcommand{\tabincell}[2]{\begin{tabular}{@{}#1@{}}#2\end{tabular}}
\scalebox{0.9}{
\begin{tabular}{c|c|p{1.2cm} p{1.2cm}| p{1.2cm} p{1.2cm}|p{1.2cm} p{1.2cm}}
\Xhline{1pt}
\multirow{2}*{\textbf{Attack}}  & \multirow{2}*{\textbf{Model \& Dataset}} & \multicolumn{2}{c|}{\textbf{Baseline}}  & \multicolumn{2}{c|}{\makecell{\textbf{$P_f$ for Fine-tuning} \\ \textbf{$P_i$ for Inference}}} & \multicolumn{2}{c}{\makecell{\textbf{$P_f$ for Fine-tuning} \\\textbf{$P_f$ for  Inference}}}\\
\cline{3-8}
~ &~& ACC & ASR  & ACC & ASR & ACC & ASR\\
\Xhline{1pt}
Trojan (WM) & \multirow{3}*{ResNet-18 (Cifar10)} &  0.900 & 0.985  & \textbf{0.810} & \textbf{0.180} & 0.790 & 0.205 \\
\cline{1-1}\cline{3-8}
Trojan (SQ) & & 0.880 & 1.000 &\textbf{0.780} & \textbf{0.040} & 0.760 & 0.065 \\
\cline{1-1}\cline{3-8}
 BadNets All-to-all & & 0.875 & 0.670 & \textbf{0.670} & \textbf{0.030} & 0.765 & 0.020 \\
\hline 
BadNets & LeNet-8 (GTSRB)&  0.960 & 0.985 & \textbf{0.905} & \textbf{0.035} & 0.875 & 0.045 \\
\hline 
Trojan (SQ)& \multirow{1}*{VGG-16 (PubFig)}&  0.955 & 1.000 & \textbf{0.870} & \textbf{0.015} & 0.815 & 0.015 \\

\Xhline{1pt}
\end{tabular}}
\caption{Evaluation of ACC and ASR with the identified solution for the five attacks in the valiation set.}
\vspace{-2em}
\label{tab:whyfinetune2}
\end{table*}

\subsection{Comparisons with Existing Works}
\label{sec:comare-other}

We compare our identified solution with some state-of-the-art defenses: Neural Cleanse with Unlearning (NC (unlearning)) \cite{wang2019neural}, Fine-pruning (FP) \cite{liu2018fine}, FLIP \cite{li2020rethinking}, and ShrinkPad-4 (SP-4) \cite{li2020rethinking}. To make a fair comparison, for all the defense methods based on fine-tuning, we set the number of clean samples as 10000. For FP, we only prune the last convolutional layer of the infected model following the same settings of the original work. We stop the pruning process when the validation accuracy is decreased by 4\% compared to the baseline ACC, as suggested in \cite{liu2018fine}. We also combine finetuning with FP, which fine-tunes the pruned model for one epoch \cite{liu2018fine}.

\begin{table}[!htbp]
\centering

\newcommand{\tabincell}[2]{\begin{tabular}{@{}#1@{}}#2\end{tabular}}
\scalebox{0.8455}{
\begin{tabular}{c|p{0.9cm} p{0.9cm}|p{0.9cm} p{0.9cm}|p{0.9cm} p{0.9cm}}
\Xhline{1pt}
\multirow{3}*{~}  & \multicolumn{2}{c|}{\textbf{Cifar10 (L2)}} & \multicolumn{2}{c|}{\textbf{Cifar10 (L0)}} & \multicolumn{2}{c}{\textbf{PubFig (WM)}} \\
\cline{2-7}

~ & ACC & ASR& ACC & ASR & ACC & ASR\\
\Xhline{1pt}
Baseline & 0.900& 0.985& 0.895& 0.990 & 0.960 & 1.000\\
\hline
\AlgName & 0.810& \textbf{0.180}& 0.825& \textbf{0.080} &0.910 & \textbf{0.010}\\
\hline 
NC (unlearning)& NA & NA & NA & NA & 0.880 & 0.025\\
\hline 
FP & 0.860& 0.990& 0.860 & 0.880 & 0.909 & 1.000\\
\hline 
FP (finetuned)& 0.895& 0.935& 0.900& 0.810 & 0.929 & 1.000\\
\hline 
FLIP & 0.900& 0.965& 0.890& 0.975 & 0.930 & 0.385\\
\hline 
SP-4 & 0.855& 0.735& 0.850& 0.985 & 0.960 & 0.995\\
\Xhline{1pt}
\end{tabular}}
\caption{ACC and ASR between our solution and prior defenses against the three attacks in the search set.}
 \vspace{-2em}
\label{tab:compare0}
\end{table}

Table \ref{tab:compare0} shows the comparison against the three attacks in the search set from the Attack Database. We see that \AlgName gets the best defense results over all the other solutions. 
Neural Cleanse fails to counter the backdoor caused by the invisible triggers as it does not consider this type of threat in its design. As a result, the outlier detector in NC cannot distinguish the target class in these attacks. 
Since its unlearning procedure is based on the detected target class label, if the detection fails to identify the target label, NC is not able to perform the unlearning procedure correctly.

We also observe that both FP and FP (finetuned) have a relatively high ASR in all three instances. This indicates the fixed criteria in FP to stop the pruning is not generalizable. FLIP and ShrinkPad-4 are not able to tackle complex triggers such as watermarks or imperceptible perturbations. This confirms the limitations of preprocessing-only solutions. To sum up, our solution from \AlgName can beat other state-of-the-art defenses on robustness and model usability.

Table \ref{tab:compare1} presents the comparisons of these solutions over the remaining attack instances in the validation set. We can draw the same conclusion as Table \ref{tab:compare0}. Particularly, we observe that NC fails to detect the backdoor caused by the BadNets All-to-all technique as it assumes there is only one target label. It does not support the case when more than one label are selected as the targets. FP and FP (fine-tuned) maintain a 100\% of ASR on the PubFig (SQ) attack, indicating that a fixed early stop criterion of 4\% accuracy drop in ACC is not effective and generalizable. This prevents the defender from monitoring the ASR to correctly determine the optimal moment of stopping the fine-pruning and balancing the security-usability trade-off. In addition, we also observe that finetuning in FP can increase the ASR in some cases (Cifar10 with WM and GTSRB with BadNets). This indicates that the finetuning operation can make the model relearn the trigger features, as discussed in \cite{li2020rethinking}. Such drawbacks make this solution impractical against backdoor attacks. Our solution can achieve the lowest ASR in most attacks while maintaining the model performance. It exhibits great comprehensiveness and effectiveness compared to other state-of-the-art solutions. 

Finally, we measure the average ACC and ASR over all eight attacks, as shown in Table \ref{tab:compare-1}. For the NC solution, since it is not able to handle the invisible or all-to-all attacks, we have to assume the defender knows the target label for backdoor removal, which is already unrealistic. From this table, we can observe that our solution gives the best defense effectiveness with the lowest average ASR of 0.053. Meanwhile, it can still maintain an acceptable average ACC of 0.831. In contrast, the most efficient method from prior works is NC, with an average ASR of 0.389 even after we make the impractical assumptions. We conclude that our solution is the optimal defense among these methods, considering all different types of attacks. 

\subsection{Mechanism Interpretation}

We use the Local Interpretable Model-Agnostic Explanations (LIME) tool~\cite{ribeiro2016why} to understand the mechanisms and effects of our solution. LIME interprets a model by perturbing its input and checking how the output changes. Specifically, it modifies a single data sample by tweaking the pixel values and observes the resulting impact on the output to determine which regions play an important role in the model predictions.
   \vspace{-1em}
\begin{figure}[!htbp]
  \centering
  \includegraphics[width=0.9\linewidth]{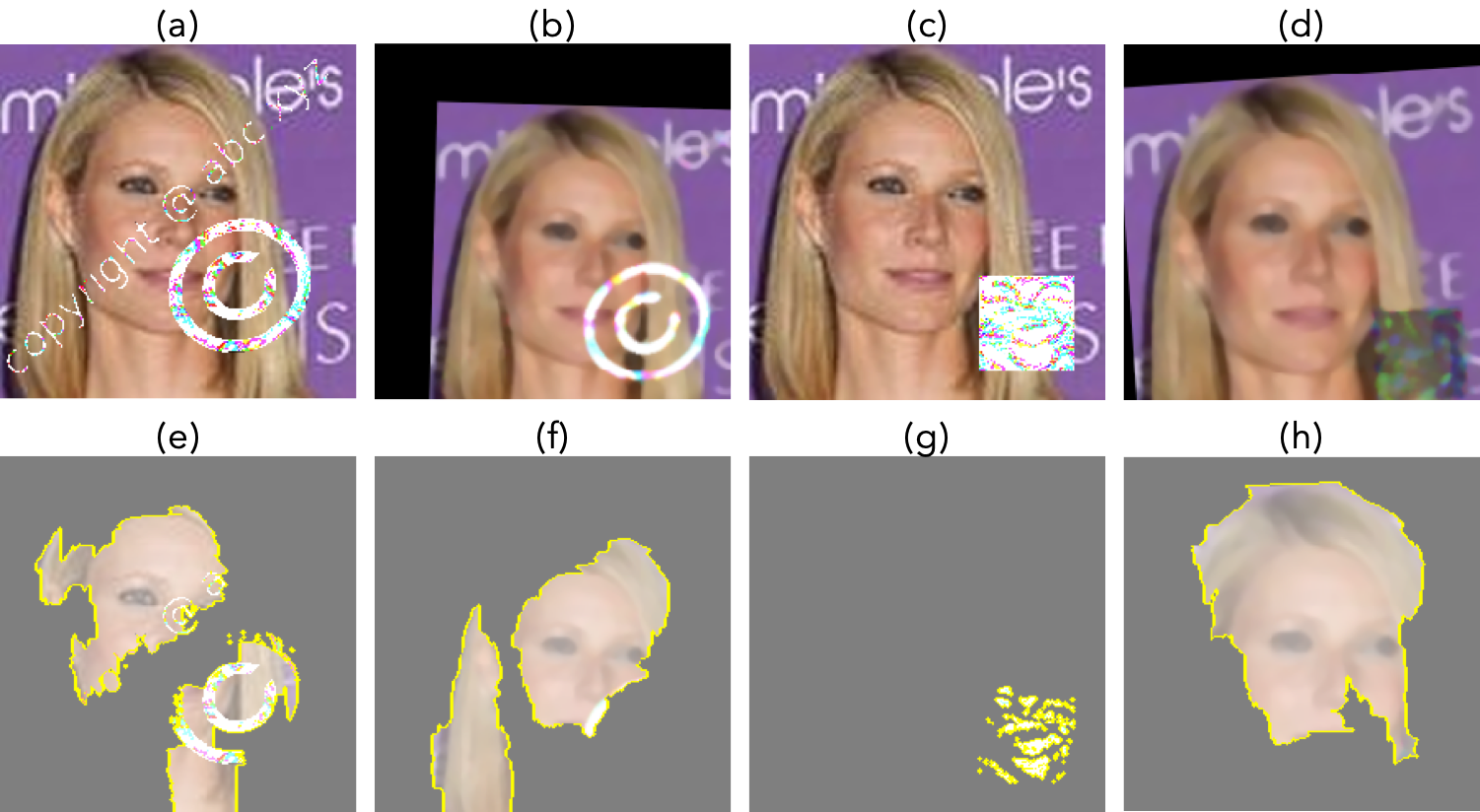}
   \vspace{-1em}
  \caption{LIME explanation for our defense solution.}
  \label{fig:lime1}

\end{figure}

In our experiment, we choose Trojan (WM) and Trojan (SQ) attacks on the PubFig dataset as examples. \figurename~\ref{fig:lime1} shows the interpretation results. The first row shows the trigger-patched samples (a and c) and their transformed output (b and d).  The second row shows the corresponding explanation results, which highlight the critical regions. We can observe that without our defense, the trigger patterns are critical to determining the classification results (e and g). After we apply our fine-tuning and inference preprocessing, the critical region now becomes the facial part of the person, which is the same as a clean image with a clean model. We conclude that our solution can successfully eliminate the trigger effects.

\section{Discussion}
\label{sec:discussion}

\subsection{Optimization of the Policies}
We use \AlgName to systematically identify the combinations of augmentation functions for the transformation policies. Although the policies can effectively mitigate backdoor impacts and preserve the model's performance from our empirical testing, they may not be the optimal solution. Our algorithm simply stacks these shortlisted functions without any optimization. Some functions may have common operations, which can be merged to make the final policy more lightweight. For instance, the operation of image resizing is adopted in many operations (e.g., DSSM, RSPA, SAT). Our policy ensemble strictly follows the operation of each selected function and performs image resizing multiple times. 
It is possible that we can combine the operations of scaling up/down from these functions into one operation, conducted before/after we execute the critical operations in all these functions. 
Also, some other operations may potentially outperform augmentation functions in this paper, such as the randomized smoothing-based approaches \cite{zhang2020backdoor,wang2020certifying}.
By including more operations, how to design an algorithm to automatically optimize and simplify the identified policies will be our future work. 

\subsection{Comprehensiveness of our Solution}
\label{sec:discuss-comp}

\begin{table*}[!htbp]
\centering
\newcommand{\tabincell}[2]{\begin{tabular}{@{}#1@{}}#2\end{tabular}}
\scalebox{0.88}{
\begin{tabular}{c|p{1.3cm} p{1.3cm}|p{1.3cm} p{1.3cm}|p{1.3cm} p{1.3cm}|p{1.3cm} p{1.3cm}|p{1.3cm} p{1.3cm}}
\Xhline{1pt}
~& \multicolumn{2}{c|}{\textbf{Cifar10 (WM)}}  & \multicolumn{2}{c|}{\textbf{Cifar10 (SQ)}} & \multicolumn{2}{c|}{\textbf{PubFig (SQ)}} & \multicolumn{2}{c|}{\textbf{GTSRB (BadNets)}}& \multicolumn{2}{c}{\textbf{Cifar10 (BadNets A2A)}}\\
\cline{2-11}
~& ACC & ASR  & ACC & ASR & ACC & ASR & ACC & ASR& ACC & ASR\\
\Xhline{1pt}
Baseline & 0.830& 1.000& 0.880& 1.000& 0.955& 1.000& 0.960& 0.985& 0.875& 0.670\\
\hline
\AlgName & 0.785&\textbf{ 0.045}& 0.780& \textbf{0.040}& 0.870& 0.015& 0.905& 0.035& 0.765& \textbf{0.020}\\
\hline 
NC (unlearning) & 0.895& 0.085& 0.910& 0.155& 0.810 & \textbf{0.010} & 0.960 & 0.190& NA & NA\\
\hline 
FP& 0.835& 0.195& 0.845& 0.235& 0.855& 1.000& 0.930& 0.020& 0.630& 0.055\\
\hline 
FP (finetuned) & 0.855& 0.650& 0.870& 0.140& 0.895& 1.000& 0.940& 0.545& 0.775& 0.055\\
\hline 
FLIP & 0.830& 0.880& 0.775& 0.090& 0.915& 0.015& 0.535& \textbf{0.005}& 0.855& \textbf{0.020}\\
\hline 
SP-4 & 0.720 & 1.000 & 0.800& 0.075& 0.940& 0.015& 0.945& 0.080& 0.625& 0.130\\
\Xhline{1pt}
\end{tabular}}
\caption{Comparing ACC and ASR between \AlgName and prior defenses on the 5 remaining attacks in the Attack Database.}
\vspace{-2em}
\label{tab:compare1}
\end{table*}

\begin{table}[!htbp]
\centering

\newcommand{\tabincell}[2]{\begin{tabular}{@{}#1@{}}#2\end{tabular}}
\scalebox{0.8455}{
\begin{tabular}{c|p{1.3cm} p{1.3cm}}
\Xhline{1pt}
~ & AvgACC & AvgASR\\
\Xhline{1pt}
Baseline & 0.907& 0.954\\
\hline
\AlgName & 0.831& \textbf{0.053}\\
\hline 
NC (unlearning) & 0.891* & 0.389*\\
\hline detect
FP & 0.841& 0.547\\
\hline 
FP (finetuned)& 0.882& 0.642\\
\hline 
FLIP & 0.828& 0.417\\
\hline 
SP-4 & 0.837& 0.502\\
\Xhline{1pt}
\end{tabular}}
\caption{Comparisons of the average ACC and ASR between our solution and prior defenses, where `*' means the results are computed by replacing `NA' with the ground truth labels.}
\vspace{-2em}
\label{tab:compare-1}
\end{table}

Although our identified solution can cover the attacks used for the search stage, as well as for validation, we cannot guarantee it is able to defeat all types of backdoor attacks. How to fundamentally solve all the backdoor attacks is still an unsolved problem. The reason behind this is that backdoor attacks can have a variety of designs and implementations. Different from adversarial examples whose scale of perturbations is strictly bounded, the pattern, size, and format of the trigger in a backdoor attack can be arbitrary. Without any restrictions on the backdoor attacks, it is challenging to have a universal solution. For instance, prior works also proposed semantic backdoor attacks, where the triggers have semantic meanings in an image (e.g., a pair of special glasses \cite{chen2017targeted}, cars with special colors \cite{bagdasaryan2020blind}). In this case, it is extremely difficult to detect the existence of such triggers as they do not have any anomaly compared to normal images. To the best of our knowledge, there are very few defense solutions considering such semantic backdoor attacks.

The goal of \AlgName is to provide an evaluation functionality for defenders to identify the defense method for certain types of backdoor attacks. By providing some examples of attack instances in this category, the defense solution is expected to mitigate other instances in the same category or their variants.  It does not guarantee the solution is able to address brand new types of attacks that are fundamentally different from the existing ones in consideration. In the future, we expect to supplement more attacks in the Attack Database, which can help produce more comprehensive solutions. 

\subsection{Possible Adaptive Attacks}

A more sophisticated adversary may try to bypass our defense solution by introducing robust backdoors and triggers that cannot be removed by our two transformation policies. This is possible but difficult as our policies involve certain random transformations on the images, preventing the adversary from deterministically figuring out the impacts of these transformations. 
To further enhance our defense, one possible solution is to identify multiple Inference Transformation Policies, and randomly apply one for each inference sample, as in \cite{raff2019barrage} to mitigate advanced adversarial examples. 

\subsection{Extension to Other Domains}
In this paper, we focus on the image classification tasks. The backdoor attacks may occur in other domains, e.g., natural language processing \cite{liu2017neural,zhai2021backdoor}, such that the image transformations cannot be applied. However, it is possible to use text augmentation techniques \cite{kobayashi2018contextual,wei2019eda} (e.g., deletion, insertion, shuffling, etc) to fine-tune the model and preprocess the inference text to defeat the corresponding backdoor attacks. Future work will focus on the design of an automatic search method for backdoor mitigation of NLP tasks.

\section{Conclusion}
\label{sec:conclusion}

This paper proposes \AlgName, a novel framework to systematically evaluate and identify defense solutions against DNN backdoor attacks. \AlgName adopts data augmentation functions to transform the infected model as well as the inference samples, the integration of which can significantly break the backdoor threats. We use this framework to produce an end-to-end solution, which is able to mitigate 8 mainstream backdoor attacks, and beat 5 state-of-the-art existing solutions from the perspectives of comprehensiveness, model usability, and robustness. 

We open-source this framework to facilitate the research of backdoor attacks for defense design and benchmarking. We will continuously maintain this framework with new emerging attacks and augmentation functions, to make the framework more comprehensive. We also expect the researchers in the AI and security communities can contribute to the development of this framework.

\section*{Acknowledge}

We thank the anonymous reviewers for their valuable comments. 
This work was supported in part by Singapore Ministry of Education AcRF Tier 1 RS02/19. 
This work was supported in part by National Key Research and Development Plan of China, 2018YFB1800301 and National Natural Science Foundation of China, 61832013.



\bibliographystyle{ACM-Reference-Format}
\bibliography{basebib.bib}


\appendix

\section*{Appendix}

\section{Augmentation Library} 
This section lists the details of all the transformation functions in our augmentation library, as shown in Table \ref{tab:auglib}. 
We try to classify these transformation functions into four main classes including the affine-transformation-based approach, the compression/quantization-based approach, noise injection/channel distortion-based approach, and the advanced transformation-based approach.

Note some of the functions in the advanced transformation-based approach are also made up of the first three approaches. 
However, since these sophisticated functions are combining multiple different approaches, we classify them together as an advanced transformation-based approach. 
Some of these functions are already deployed to mitigate the adversarial examples with a high level of image content changing while still maintaining high ACCs. 
It is necessary to use them as potential candidates in our evaluation framework.

\section{Algorithms and Parameters}

We present the details of the augmentation candidates used in the policies of \AlgName. The hyperparameters we adopted for each augmentation are in Table \ref{tab:Hyper}.

\begin{table}[!htbp]
\small
\centering
\newcommand{\tabincell}[2]{\begin{tabular}{@{}#1@{}}#2\end{tabular}}
\begin{tabular}{c c c}
\Xhline{1pt}
\textbf{Notation} & \textbf{Meaning} & \textbf{Value}\\
\Xhline{1pt}
$\delta$ & distortion limit of the Optical Distortion & 0.5\\
\hline 
$\gamma_1$ & GCSM's gamma value & 0.6\\
\hline 
$\gamma_2$ & GESM's gamma value & 2.6\\
\hline 
$\sigma$ & scale limit of the RSPA & 1.3\\
\hline 
$T$ & translation limit of the SAT & 0.16\\
\hline 
$S$ & sacaling limit of the SAT & 0.16\\
\hline 
$R$ & rotation limit of the SAT & 4\\
\Xhline{1pt}
\end{tabular}
\caption{Hyperparameters' settings used during the Preprocessing in this paper.}
\label{tab:Hyper}
\end{table}

\subsection{Optical Distortion}
Different from \cite{liu2010pincushion}, the Optical Distortion we upgraded and utilized in the \AlgName is based on assigning a random distortion value chosen from a uniform distribution of the distortion limit. This random process can distort each sample on a different scale for a different time, thus better help the infected model better adapt to the remapping distortions. The details of the Random Pincushion Distortion we proposed and improved in the \AlgName are explained in Algorithm \ref{algo:OD}. The random pincushion distortion can be interpreted into three phases. For starters, we acquire a random distortion value, $\delta_k$,  from a uniform distribution between $-\delta$ to $0$. Using this randomly sampled $\delta_k$, we can acquire two pincushion maps for horizontal and vertical indexes, respectively. Finally, by broadcasting those two maps for each pixel, we can output the result. During the experiment, we set the $\delta$ as 0.5 based on experimental analysis.

\SetKwInput{KwParam}{Parameters}
\begin{algorithm}[]
\small
    \caption{\emph{Random Pincushion Distortion}}
    \label{algo:OD}
    \SetNoFillComment
    \KwIn{original image $I\in \mathbb{R}^{h\times w}$}
    \KwOut{distorted image $I'\in \mathbb{R}^{h\times w}$}
    \KwParam{distortion limit $\delta$;}
    \BlankLine
      
      \tcc{1.Acquire distortion parameter $\delta_k$}
      $\delta_k\sim\mathcal{U}(-\delta, 0)$\;
      \tcc{2.Acquire Distortion Maps}
      $c_x=\left\lfloor(w/2)\right\rfloor$, $c_y=\left\lfloor(h/2)\right\rfloor$\;
      $P_{set}=\{(m, n)\in\{(0, ..., w)\times(0, ..., h)\}\}$\;
      \For{$(u, v)$ in $P_{set}\backslash\{(m, n)\}$}{
          $map_x(u,v) = ((u-c_x)\times(1+k))+c_x$\;
          $map_y(u,v) = ((v-c_y)\times(1+k))+c_y$\;
      }
      \tcc{3.Remapping $I$ to $I'$}
      \For{$(u, v)$ in $P_{set}\backslash\{(m, n)\}$}{
      $I'(u,v)=I(map_x(u,v),map_y(u,v))$\;
      }
      \Return $I'$\;
\end{algorithm}

\subsection{Gamma Compression and Extension}

\begin{figure*}[!htbp]
  \centering
  \includegraphics[width=0.99\textwidth]{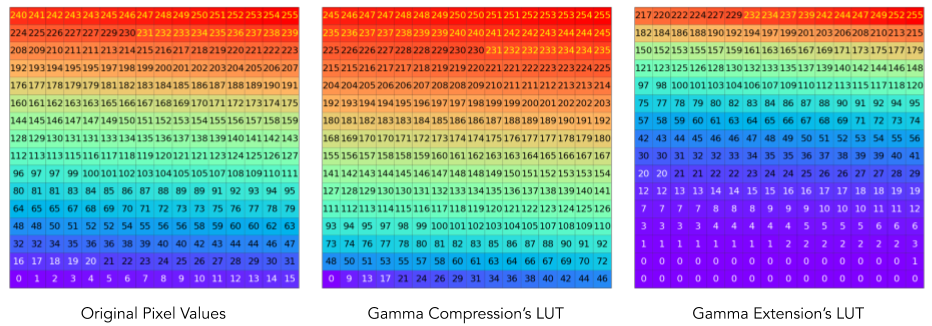}
  \caption{Different Gamma Look-Up Tables (LUTs) used in the Median Filters set: The left is the original pixel values, ranging from 0 to 255; the Gamma Compression uses a Gamma value of 0.6, which lead to the LUT shown in the middle; the Gamma Extention uses a Gamma value of 2.6, which lead the left LUT.}
  \label{fig:gamma}
\end{figure*}

Inspired by the previous work \cite{kumari2014single}, the Gamma Compression and the Gamma Extension are fine-tuned and used in the median filters set to merging pixels' values and enhance the effects of the median filters, namely the GCSM and GESM. The Gamma value of the Gamma Compression procedure is set to 0.6, which acquires a Look-Up Table shown in the middle of Figure \ref{fig:gamma}. 
As demonstrated that larger values from the original pixels range (the left part of Figure \ref{fig:gamma}) are mapping with a larger value close to the maximum value (255), thus helps larger values to bend in. As a result, the median filter can work more efficiently to smoothen pixels of low value. Vice versa, with a Gamma value set to 2.6, we can use the help of the Gamma Extension to merge small values, thus better smoothen large pixels. We summarize the Gamma Compression and Extension as a single function shown in Algorithm \ref{algo:GT}. As demonstrated, the Gamma Transformation we used here in the experiment can be interpreted as two functional parts. First, we acquire the LUT based on the Gamma value, $\gamma$. Then, the output image can be obtained by using the value of the corresponding position in the LUT to replace the original pixel value. The function with a Gamma value larger than 1 conducts extension, and a Gamma value smaller than 1 performs compression. We chose 0.6 and 2.6 as the Gamma values for the compression and extension based on experimental results.

\SetKwInput{KwParam}{Parameters}
\begin{algorithm}[h]
\small
    \caption{\emph{Gamma Transformation}}
    \label{algo:GT}
    \SetNoFillComment
    \KwIn{original image $I\in \mathbb{R}^{h\times w}$}
    \KwOut{transformed image $I'\in \mathbb{R}^{h\times w}$}
    \KwParam{Gamma Value $\gamma$;}
    \BlankLine
      
      \tcc{1.Acquire LUT}
      $T = range(0:255)^{16\times16}$\;
      $LUT = (T/255)^{\gamma}\times 255$\;   
      \tcc{2.Assigning New Values}
      $P_{set}=\{(m, n)\in\{(0, ..., w)\times(0, ..., h)\}\}$\;
      \For{$(u, v)$ in $P_{set}\backslash\{(m, n)\}$}{
          $(x,y)=where(T==I(u,v))$\;
          $I'(u,v)=LUT(x,y)$\;
      }
      \Return $I'$\;
\end{algorithm}


\subsection{Random Sized Padding Affine~(RSPA)}

We use Random Sized Padding Affine~(RSPA)\cite{qiu2020fencebox} as a tool to help the infected model better adapt to affine transformations. The details of the proposed preprocessing function are explained in \ref{algo:RSPA}. The $\sigma$ we used in the experiment is set to 1.3 to downscale the input image in a range of range (0.8,1). The whole process of the RSPA can be interpreted as three functional parts. First, the algorithm acquires random parameters for the scaling and the padding. This includes after-padding size, $Len_{max}$; resizing size, $Len$; the number of pixels to pad to reach the after-padding size, $l_{rem}$; and padding coordinates, $(x_1,x_2)$ and $(y_1,y_2)$. Padding the resized image using the padding coordinates to $(Len_{max}, Len_{max})$, we can acquire a black canvas patched with the resized original input. By resizing the image back to the original size, we can acquire the final result. In the experiment, resizing the image from 0.8 to 1 times smaller can best help the infected model adapt to the transformation.

\SetKwInput{KwParam}{Parameters}
\begin{algorithm}[!htbp]
\small
    \caption{\emph{RSPA}}
    \label{algo:RSPA}
    \SetNoFillComment
    \KwIn{original image $I\in \mathbb{R}^{l\times l}$}
    \KwOut{distorted image $I'\in \mathbb{R}^{l\times l}$}
    \KwParam{scale limit $\sigma$;}
    \BlankLine
      
      \tcc{1.Acquire random parameter}
      $Len_{max}=\left \lfloor (l \times \sigma) \right \rfloor$\;
      $Len\sim\left\lfloor\mathcal{U}(l, Len_{max})\right\rfloor$\;
      $l_{rem}=Len_{max}-Len$\;
      $x_1\sim\left\lfloor\mathcal{U}(0, l_{rem})\right\rfloor$,
      $y_1\sim\left\lfloor\mathcal{U}(0, l_{rem})\right\rfloor$\;
      $x_2 = l_{rem}-x_1$,
      $y_2 = l_{rem}-y_1$\;
      \tcc{2.Padding to $Len_{max}$}
      $I' = reshape(I)$ s.t. $I'\in \mathbb{R}^{Len \times Len}$\;
      $I' = pad(I',((x_1,x_2),(y_1,y_2)),value=0)$ \qquad \qquad \ 
      s.t. $I'\in \mathbb{R}^{Len_{max} \times Len_{max}}$\;
      \tcc{3.Reshape $I'$ to the size of $I$}
      $I'= \texttt{reshape}(I')$ s.t. $I'\in \mathbb{R}^{l\times l}$\;
      \Return $I'$\;
\end{algorithm}

\subsection{Stochastic Affine Transformation}

We adopt the Stochastic Affine Transformation (SAT) \cite{zeng2020data} 
in \AlgName. The parameters of the SAT in the Algorithm \ref{algo:SAT} are the same with \cite{zeng2020data}: $T$, 0.16, $S$, 0.16, and $R$, 4. 


\SetKwInput{KwParam}{Parameters}
\begin{algorithm}[!htbp]
    \scriptsize
    \caption{{SAT}}
    \label{algo:SAT}
    \SetNoFillComment
    \KwIn{original image $I\in \mathbb{R}^{h\times w}$}
    \KwOut{transformed image $I'\in \mathbb{R}^{h\times w}$}
    \KwParam{translation limit $T$; scaling limit $S$, rotation limit $R$.}
    \BlankLine
      $I^{'}=O^{h\times w}$\;
      \tcc{1.Translation}
      $\delta_x\sim\mathcal{U}(-T, T)$\; 
      $\delta_y\sim\mathcal{U}(-T, T)$\;
      $\Delta_x = \delta_x \times w$\; 
      $\Delta_y = \delta_y \times h$\;
      \If{$(x+\Delta_x \in (0, w)) \wedge (y+\Delta_y \in (0,h))$}{
          $I'(x,y)= I(x+\Delta_x,y+\Delta_y)$\; }
      \tcc{2.Rotation}
      $\delta_r\sim\mathcal{U}(-R, R)$\;
      $\Delta_r = \delta_r \times \pi/180$\;
      \For{$(x_i,y_j)$ in $\left \{ (x,y)|x\in(0,w),y\in(0,h)\right \}$}{
      $x_{i}^{'}=-(x_i-\left \lfloor w/2 \right \rfloor)\times sin(\Delta_r)+(y_j-\left \lfloor h/2 \right \rfloor)\times cos(\Delta_r)$\;
      $y_{j}^{'}=(x_i-\left \lfloor w/2 \right \rfloor)\times cos(\Delta_r)+(y_j-\left \lfloor h/2 \right \rfloor)\times sin(\Delta_r)$\;
      $x_{i}^{'}=\left \lfloor x_{i}^{'}+\left \lfloor w/2 \right \rfloor\right \rfloor$\;
      $y_{j}^{'}=\left \lfloor y_{j}^{'}+\left \lfloor h/2 \right \rfloor\right \rfloor$\;
      \If{$(x_{i}^{'} \in (0, w)) \wedge (y_{j}^{'} \in (0,h))$}{
          $I'(x_i,y_j)= I(x_{i}^{'},y_{j}^{'})$\; }
      }
      \tcc{3.Scaling}
      $\delta_s\sim\mathcal{U}(1-S, 1+S)$\;
      $h_{new} = \delta_s \times h$\;
      $w_{new} = \delta_s \times w$\;
      $I'= \texttt{reshape}(I',(h_{new},w_{new}))$\;
      \If{$\delta_s>1$}{
          $I'(x,y)= cropping(I',(h,w))$\; }
      \If{$\delta_s<1$}{
          $I'(x,y)= padding(I',(h,w))$\; }
      
      \Return $I'$\;
\end{algorithm}


\begin{table*}[!htbp]
\centering
\newcommand{\tabincell}[2]{\begin{tabular}{@{}#1@{}}#2\end{tabular}}
\scalebox{0.8455}{
\begin{tabular}{p{0.5cm} p{3.6cm}|p{0.5cm} p{3.6cm}|p{0.5cm} p{3.6cm}|p{0.5cm} p{3.6cm}}
\Xhline{1pt}
\multicolumn{2}{c|}{\makecell{\textbf{Affine-Transformation} \\ \textbf{Based}}} & \multicolumn{2}{c|}{\makecell{\textbf{Compression/Quantization} \\ \textbf{Based}}} & \multicolumn{2}{c|}{\makecell{\textbf{Noise Injection} \\ \textbf{/Channel Distortion Based}}} & \multicolumn{2}{c}{\makecell{\textbf{Advanced} \\ \textbf{Transformation Based}}} \\
\Xhline{1pt}

Index & Name & Index & Name & Index & Name & Index & Name\\
\Xhline{1pt}
1 & VerticalFlip & 23 & Normalize & 39 & Blur & 66 & SHIELD \\
2 & HorizontalFlip & 24 & DSSM & 40 & RandomGamma & 67 & PixelDeflection\\
3 & Flip & 25 & GCSM & 41 & RandomBrightness & 68 & Bit-depth Reduction \\
4 & Transpose & 26 & GESM & 42 & RandomContrast & 69 & RSPA \\
5 & RandomCrop & 27 & MedianBlur & 43 & MotionBlur & 70 & SAT \\
6 & RandomRotate90 & 28 & CLAHE & 44 & GaussianBlur & 71 & Feature Distillation \\
7 & Rotate & 29 & JpegCompression & 45 & GaussNoise &   &  \\
8 & ShiftScaleRotate & 30 & ImageCompression & 46 & GlassBlur &   &  \\
9 & CenterCrop & 31 & Downscale & 47 & ChannelShuffle &   &  \\
10 & OD & 32 & MultiplicativeNoise & 48 & InvertImg &   &  \\
11 & GridDistortion & 33 & FancyPCA & 49 & ToGray &   &  \\
12 & ElasticTransform & 34 & Posterize & 50 & ToSepia &   &  \\
13 & RandomGridShuffle & 35 & LowPassFilter & 51 & CoarseDropout &   &  \\
14 & Cutout & 36 & RandomWebP & 52 & RGBShift &   &  \\
15 & Crop & 37 & HighPassFilter & 53 & RandomBrightnessContrast &   &  \\
16 & RandomScale & 38 & RandomValueFit & 54 & RandomCropNearBBox &   &  \\
17 & LongestMaxSize &   &   & 55 & RandomSizedBBoxSafeCrop &   &  \\
18 & SmallestMaxSize &   &   & 56 & RandomSnow &   &  \\
19 & Resize &   &   & 57 & RandomRain &   &  \\
20 & RandomSizedCrop &   &   & 58 & RandomFog &   &  \\
21 & RandomResizedCrop &   &   & 59 & RandomSunFlare &   &  \\
22 & GridDropout &   &   & 60 & RandomShadow &   &  \\
  &   &   &   & 61 & ChannelDropout &   &  \\
  &   &   &   & 62 & ISONoise &   &  \\
  &   &   &   & 63 & SolarizeEqualize &   &  \\
  &   &   &   & 64 & Equalize &   &  \\
  &   &   &   & 65 & ColorJitter &   &  \\

\Xhline{1pt}
\end{tabular}}
\caption{Augmentation Library used in this paper: 4 main class with 71 transformation functions in total.}
\label{tab:auglib}
\end{table*}

\end{document}